\documentclass[11pt]{article}
\usepackage[T1]{fontenc}
\usepackage[margin=1in]{geometry}
\usepackage{amsmath,amssymb,amsthm,bm,booktabs,graphicx,xcolor}
\usepackage[labelsep=period,font=small]{caption}
\usepackage{titlesec}
\titleformat{\section}{\normalfont\bfseries\large}{\thesection.}{0.6em}{\MakeUppercase}
\titleformat{\subsection}{\normalfont\bfseries}{\Alph{subsection}.}{0.6em}{}
\usepackage[hidelinks]{hyperref}
\graphicspath{{figures/}}

\newtheorem{theorem}{Theorem}
\newtheorem{proposition}{Proposition}
\newtheorem{lemma}{Lemma}
\theoremstyle{definition}
\newtheorem{definition}{Definition}
\theoremstyle{remark}
\newtheorem{remark}{Remark}

\newcommand{\vx}{\bm{x}}
\newcommand{\vu}{\bm{u}}
\newcommand{\va}{\bm{a}}
\newcommand{\vf}{\bm{f}}
\newcommand{\ve}{\bm{e}}
\newcommand{\vX}{\bm{X}}
\newcommand{\dd}{\,\mathrm{d}}
\newcommand{\tr}{\operatorname{tr}}
\newcommand{\Lap}{\nabla^2}
\newcommand{\bmsection}[1]{\par\medskip\noindent\textbf{#1.}\enspace\ignorespaces}
\newcommand{\tablenote}[1]{\par\smallskip\begin{minipage}{0.86\textwidth}\footnotesize #1\end{minipage}}

\title{When is defocus a diffusion? Exact differential-defocus constraints in wave optics}
\author{Junjie Luo\textsuperscript{*}\\[6pt]
\small Elmore Family School of Electrical and Computer Engineering, Purdue University\\
\small\textsuperscript{*}luo330@purdue.edu}
\date{}

\begin{document}
\maketitle

\begin{abstract}
\noindent Depth from differential defocus (DfDD) recovers depth from small changes of focus, aperture or object
position through local constraints in which defocus acts as a diffusion of the image. In geometric optics these
constraints are exact only for a Gaussian aperture. We study them in scalar wave optics, where the optical transfer
function is an integral over the overlap of two displaced copies of the pupil. For a Gaussian pupil every DfDD
constraint holds in closed form: focus, sensor-distance and lateral-motion cues are unaffected by diffraction, while
axial motion, aperture changes and wavelength changes acquire a diffraction term, which folds the depth map of coupled
optical differentiation near focus. Conversely, we prove that defocus is an exact diffusion, possibly anisotropic and
with a drift, if and only if the pupil has a Gaussian amplitude and a phase of degree at most two. Hard-edged apertures
never qualify, whatever their phase.
\end{abstract}

\section{Introduction}
\label{sec:intro}

A defocused image is, to a good approximation, the focused image diffused: when the blur grows slightly, the image
changes by a multiple of its own Laplacian, as in the heat equation~\cite{favaro2007}. Depth from differential defocus
(DfDD) turns this observation into a depth sensor~\cite{alexander2019}. Two or three images taken with slightly
different focus, aperture or object position give the change of the image, and the ratio of that change to the
Laplacian gives the defocus, and hence the depth, at every pixel in closed form. Focal flow combines defocus with
motion~\cite{alexander2018}, focal track changes the optical power of a deformable lens~\cite{guo2017}, focal split
records two focus states in one exposure~\cite{luo2025fs}, and coupled optical differentiation (COD) takes depth from
the ratio of an aperture derivative to a focus derivative~\cite{luo2025cod}.

These constraints were derived in geometric optics, where the point-spread function (PSF) is a scaled copy of the
aperture. Alexander showed that a local constraint of this kind holds exactly only when the aperture has a Gaussian
transmission profile~\cite{alexander2019}; for other apertures it is an approximation whose error grows with the blur.
Real images, however, are formed by diffraction as well as by defocus, and diffraction matters most in the small,
low-power cameras that motivate DfDD. Two questions follow. Does a Gaussian aperture keep the DfDD constraints exact
once diffraction is included, and what happens to the cues that change the diffraction blur itself, such as aperture
changes or axial motion? And does wave optics, which allows complex pupils with phase masks, admit other pupils with an
exact local constraint?

\begin{figure}[t]
\centering
\includegraphics[width=\textwidth]{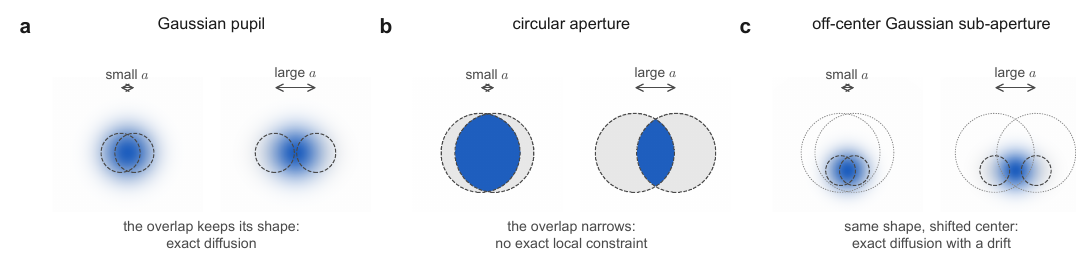}
\caption{The overlap of two copies of the pupil, displaced by $\va=\lambda s\vf$, decides whether defocus is a
diffusion. Dashed: the two copies (half maximum for Gaussian pupils); blue: their overlap, which transmits frequency
$\vf$. (a)~For a Gaussian pupil the overlap has the same shape at every displacement, and defocus is an exact diffusion.
(b)~For a circular aperture the overlap narrows as the displacement grows; no local constraint is exact. (c)~An
off-center Gaussian sub-aperture (dotted: the full pupil) keeps the shape of its overlap, whose off-axis center adds a
drift proportional to the defocus.}
\label{fig:kinds}
\end{figure}

We answer both questions in scalar, paraxial wave optics. Section~\ref{sec:model} writes the optical transfer function
(OTF) at spatial frequency $\vf$ as an integral over the overlap of two copies of the pupil displaced by $\lambda s\vf$,
a classical picture of the OTF~\cite{goodman2005,brenner1983}, and shows that a local constraint exists if and only if
the logarithmic derivative of the OTF is a polynomial of degree two in $\vf$. For a Gaussian pupil the OTF stays
Gaussian at every defocus, and its variance is the sum of a geometric and a diffraction variance
(Sec.~\ref{sec:gauss}). Focus, sensor-distance and lateral-motion cues then remain exact and independent of
diffraction, while axial motion, aperture changes and wavelength changes acquire a term proportional to the diffraction
variance. For COD the same term makes two depths on opposite sides of focus give the same measurement
(Sec.~\ref{sec:cod}). Our main result is a converse (Sec.~\ref{sec:unique}): defocus is an exact diffusion, possibly
anisotropic and with a drift, if and only if the pupil has a Gaussian amplitude and a phase of degree at most two.
Figure~\ref{fig:kinds} shows why. For a Gaussian pupil the overlap of the two copies keeps its shape at every
displacement, for a circular aperture it narrows, and for an off-center Gaussian sub-aperture it keeps its shape but
not its position, which produces the drift that dual-pixel sensors measure~\cite{garg2019,punnappurath2020}. A companion
paper~\cite{luo2026dfdd} studies what happens instead with the circular apertures of real lenses, where the diffusion
coefficient falls with spatial frequency and depth estimates become texture-dependent.

\section{DfDD in geometric optics}
\label{sec:background}

Consider a fronto-parallel plane with texture $T$ at depth $Z$ and lateral position $\vX$, imaged by a thin lens of
optical power $\rho$ onto a sensor at distance $s$~\cite{luo2025fs,luo2025cod}. The image is the convolution of the
pinhole image $P$ with the PSF $k$,
\begin{equation}
  I(\vx)=k(\vx)*P(\vx),\qquad P(\vx)=T\Big(-\frac Zs\vx-\vX\Big).
  \label{eq:image}
\end{equation}
In geometric optics the PSF is a scaled copy of the aperture transmittance profile $\kappa$,
\begin{equation}
  k(\vx)=\frac1{\sigma^2}\,\kappa\Big(\frac{\vx}{\sigma}\Big),\qquad \sigma=As\psi,\qquad \psi=\frac1Z-\rho+\frac1s,
  \label{eq:sigma}
\end{equation}
with the aperture scale $A$, the defocus level $\sigma$ and the defocus $\psi$ in diopters. For a Gaussian profile of
unit variance, $I$ obeys the heat equation $\partial I/\partial\sigma=\sigma\Lap I$. Since $\partial\sigma/\partial\rho
=-As$, a focus change gives
\begin{equation}
  I_\rho=-As\,\sigma\,\Lap I=-A^2s^2\psi\,\Lap I ,
  \label{eq:dbc}
\end{equation}
so the ratio $I_\rho/\Lap I$ gives $\psi$, and $\psi$ gives the depth. Allowing every parameter to change with time
gives the constraint of focal flow and focal track~\cite{alexander2018,guo2017,alexander2019},
\begin{equation}
  \dot I=\tfrac sZ\,\dot\vX\cdot\nabla I+\big(\tfrac{\dot Z}{Z}-\tfrac{\dot s}{s}\big)\,\vx\cdot\nabla I
  +\sigma\big[\dot\sigma+\big(\tfrac{\dot Z}{Z}-\tfrac{\dot s}{s}\big)\sigma\big]\Lap I ,
  \label{eq:geo}
\end{equation}
where dots are time derivatives. COD measures both $I_\rho$ and the aperture derivative, $I_A=(\sigma^2/A)\Lap I$; the
ratio $q=A\,I_A/I_\rho=-\psi$ then gives depth for any aperture profile~\cite{luo2025cod}. For non-Gaussian apertures,
\eqref{eq:dbc} and~\eqref{eq:geo} hold only in the limit of small blur.

\section{Defocus in wave optics}
\label{sec:model}

We use scalar, paraxial wave optics with quasi-monochromatic light of wavelength $\lambda$ and an incoherently lit
target. Let $p(\vu)$ be the complex pupil function at position $\vu$ in the pupil. The OTF $H$, the Fourier transform of
the PSF, is then~\cite{goodman2005}
\begin{equation}
\begin{aligned}
  H(\vf;\psi)&=\int O_{\va}(\vu)\,e^{-i2\pi\psi\,\vu\cdot\va/\lambda}\dd\vu,\\
  O_{\va}(\vu)&=\frac{p(\vu-\tfrac{\va}{2})\,p^*(\vu+\tfrac{\va}{2})}{\|p\|^2},\qquad \va=\lambda s\vf .
\end{aligned}
  \label{eq:otf}
\end{equation}
The function $O_{\va}$ is the overlap of two copies of the pupil displaced by $\pm\va/2$: image detail at frequency
$\vf$ is formed by light from pairs of pupil points $\va$ apart. Defocus adds a phase that varies linearly across the
overlap, so $H$ is the Fourier transform of the overlap evaluated at $\psi\va/\lambda$. Brenner et
al.~\cite{brenner1983} express the same fact through the ambiguity function of the pupil. As $\lambda\to0$ at fixed
$\psi s\vf$, the overlap becomes the pupil intensity $|p|^2/\|p\|^2$, and~\eqref{eq:otf} reduces to the geometric PSF
of~\eqref{eq:sigma}.

When only the optics change and the scene is static, $\hat I=H\hat P$, and the derivative of the image with respect to
any optical parameter $t$ is a filtered copy of the image,
\begin{equation}
  \widehat{I_t}(\vf)=\partial_t\log H(\vf)\;\hat I(\vf).
  \label{eq:spectral}
\end{equation}
This holds for changes of $\rho$, $A$ and $\lambda$, and for changes of $s$ after correcting the magnification. A local
constraint writes $I_t$ as a fixed combination of the image and its first and second derivatives, which multiply the
spectrum by polynomials of degree one and two in $\vf$. By~\eqref{eq:spectral}, such a constraint holds for every
texture if and only if $\partial_t\log H$ is a polynomial of degree at most two in $\vf$; since $H(0)=1$ it has no
constant term. The heat equation~\eqref{eq:dbc} is the case $\partial_\rho\log H\propto|\vf|^2$.

\section{Gaussian pupils}
\label{sec:gauss}

We state the results for an anisotropic Gaussian pupil, $p(\vu)=\exp(-\vu^\top\Gamma^{-1}\vu/(4A^2))$, whose intensity has
covariance $A^2\Gamma$; the isotropic case is $\Gamma=I$.

\begin{lemma}[Gaussian pupil]
\label{lem:gauss}
For this pupil,
\begin{equation}
  H(\vf;\psi)=\exp\!\big(-2\pi^2\,\vf^\top C\vf\big),\qquad C=\sigma^2\Gamma+\sigma_d^2\Gamma^{-1},\qquad
  \sigma=As\psi,\qquad \sigma_d=\frac{\lambda s}{4\pi A}.
  \label{eq:gauss_otf}
\end{equation}
\end{lemma}
\begin{proof}
$p(\vu-\tfrac{\va}{2})\,p(\vu+\tfrac{\va}{2})=\exp(-\va^\top\Gamma^{-1}\va/(8A^2))\,\exp(-\vu^\top\Gamma^{-1}\vu/(2A^2))$, so
the shape of the overlap does not depend on $\va$ (Fig.~\ref{fig:kinds}a). Its Fourier transform at $\psi\va/\lambda$,
normalized by $\|p\|^2$, is $\exp(-\va^\top\Gamma^{-1}\va/(8A^2))\exp(-2\pi^2A^2\psi^2\va^\top\Gamma\va/\lambda^2)$.
Substituting $\va=\lambda s\vf$ gives~\eqref{eq:gauss_otf}.
\end{proof}

The PSF is therefore an exact Gaussian at every defocus, as for a Gaussian beam~\cite{siegman1986}, with covariance
$\sigma^2\Gamma+\sigma_d^2\Gamma^{-1}$: the geometric blur plus a diffraction blur. A separable model, a geometric
Gaussian convolved with a diffraction kernel, is thus exact for Gaussian pupils, and Theorem~\ref{thm:unique} below
implies that it is exact only for them. The diffraction blur $\sigma_d$ is a standard deviation; a hard aperture of
radius $R=2A$ has the same second moment, and then $\sigma_d=\lambda N/\pi$ with the f-number $N=s/(2R)$, about a
quarter of the Airy radius $1.22\lambda N$.

The diffraction blur does not depend on the focus, so the focus derivative ignores it. It does depend on the aperture,
on the wavelength and, through the magnification, on the depth of the target. Cues that change these quantities pick up
a new term.

\begin{proposition}[DfDD constraint in wave optics]
\label{prop:wave}
Under Lemma~\ref{lem:gauss}, for small changes of $\rho$, $A$, $s$, $\vX$, $Z$ and $\lambda$,
\begin{equation}
\begin{aligned}
  \dot I={}&\tfrac sZ\,\dot\vX\cdot\nabla I+\big(\tfrac{\dot Z}{Z}-\tfrac{\dot s}{s}\big)\,\vx\cdot\nabla I
  +\sigma\big[\dot\sigma+\big(\tfrac{\dot Z}{Z}-\tfrac{\dot s}{s}\big)\sigma\big]\tr(\Gamma\nabla\nabla^\top)I\\
  &+\sigma_d^2\big(\tfrac{\dot Z}{Z}-\tfrac{\dot A}{A}+\tfrac{\dot\lambda}{\lambda}\big)\tr(\Gamma^{-1}\nabla\nabla^\top)I .
\end{aligned}
  \label{eq:wave}
\end{equation}
\end{proposition}
\begin{proof}
Write $I=G_C*P$, with $G_C$ the Gaussian of covariance $C$ and $P$ from~\eqref{eq:image}.
(a) Heat equation: $\partial I/\partial C_{ij}=\tfrac12\partial_i\partial_jI$, which contributes
$\tfrac12\tr(\dot C\nabla\nabla^\top)I$.
(b) Pinhole transport: $\dot P=\tfrac sZ\dot\vX\cdot\nabla P+(\tfrac{\dot Z}{Z}-\tfrac{\dot s}{s})\,\vx\cdot\nabla P$.
(c) $G_C*(\vx\cdot\nabla P)=\vx\cdot\nabla I+\tr(C\nabla\nabla^\top)I$, because $(\vx-\bm y)\,G_C(\vx-\bm y)=-C\nabla_{\vx}
G_C(\vx-\bm y)$.
(d) $\tfrac12\dot C=\sigma\dot\sigma\,\Gamma+\sigma_d\dot\sigma_d\,\Gamma^{-1}$ with
$\dot\sigma_d=\sigma_d(\dot s/s-\dot A/A+\dot\lambda/\lambda)$.
Collecting the $\Gamma^{-1}$ terms of (a) and (c), the $\dot s/s$ part of $\dot\sigma_d$ cancels against the $-\dot s/s$
of the magnification term, which leaves the second line of~\eqref{eq:wave}. The $\Gamma$ terms give the first line.
\end{proof}

The first line of~\eqref{eq:wave} is the geometric constraint~\eqref{eq:geo}; the second line is new.
Table~\ref{tab:cues} lists the change for each cue. Focus, sensor-distance and lateral-motion cues are free of
diffraction, and since none of their coefficients depends on $\lambda$, they stay exact for any spectrum and any
texture color if the lens has no chromatic aberration. Axial motion, aperture changes and wavelength changes pick up a
term $\sigma_d^2$. For broadband light this term becomes $\int S(\lambda)\,\sigma_d(\lambda)^2\,\Lap I_\lambda\dd\lambda$,
which equals $\overline{\sigma_d^2}\,\Lap I$ only for a gray texture whose color channels are blurred alike. For an
anisotropic pupil the new term involves the dual operator $\tr(\Gamma^{-1}\nabla\nabla^\top)$, so the constraint gains
a basis image that the geometric constraint lacks.

\begin{table}[t]
\centering
\caption{Which DfDD cues stay exact in wave optics for a Gaussian pupil ($\Gamma=I$)}
\label{tab:cues}
\begin{tabular}{lll}
\toprule
Changed parameter & Extra term & Exact? \\
\midrule
optical power $\rho$ (focus) & none & yes \\
sensor distance $s$ & none & yes \\
lateral position $X,Y$ & none & yes \\
depth $Z$ (axial motion) & $+\sigma_d^2\,\dot Z/Z$ & no \\
aperture $A$ & $-\sigma_d^2\,\dot A/A$ & no \\
wavelength $\lambda$ & $+\sigma_d^2\,\dot\lambda/\lambda$ & no \\
\bottomrule
\end{tabular}
\tablenote{Extra term: the change of the coefficient of $\Lap I$ in Eq.~\eqref{eq:wave} caused by diffraction.}
\end{table}

\paragraph{Magnification-corrected frame.}
For $\tilde I(\vx)=I(\tfrac{s}{s_0}\vx)$ with a fixed reference distance $s_0$, the covariance is
$\tilde C=\beta^2\Gamma+\beta_d^2\Gamma^{-1}$ with $\beta=As_0\psi$ and $\beta_d=\lambda s_0/(4\pi A)$, and
\begin{equation}
  \dot{\tilde I}=\frac{s_0}{Z}\,\dot\vX\cdot\nabla\tilde I+\frac{\dot Z}{Z}(\vx\cdot\nabla\tilde I)
  +\beta\Big(\dot\beta+\frac{\dot Z}{Z}\beta\Big)\tr(\Gamma\nabla\nabla^\top)\tilde I
  +\beta_d^2\Big(\frac{\dot Z}{Z}-\frac{\dot A}{A}+\frac{\dot\lambda}{\lambda}\Big)\tr(\Gamma^{-1}\nabla\nabla^\top)\tilde I .
  \label{eq:wave_corr}
\end{equation}
The rescaling freezes the diffraction kernel for every rate except $\dot A$ and $\dot\lambda$, and the remaining $\dot
Z$ term comes from the target magnification acting on this frozen kernel.

\paragraph{Numerical check.}
We rendered exact images of a cosine-sum texture through an isotropic Gaussian pupil with the closed form of
Lemma~\ref{lem:gauss}: $A=0.25$\,mm, $s=12.1$\,mm, focus at 0.85\,m and $Z=0.70$\,m, so $\sigma=0.76\,\mu$m and
$\sigma_d=2.1\,\mu$m, and took time derivatives by central differences. Table~\ref{tab:check} gives the relative residual
of~\eqref{eq:wave} with and without the diffraction term. Without it, the residual is 0.13\% for axial motion, 115\%
for aperture changes and 100\% for wavelength changes; with it, every residual is below $3\times10^{-8}$. For broadband
light, the spectrally averaged coefficient leaves a residual of $9\times10^{-6}$.

\begin{table}[t]
\centering
\caption{Relative rms residual of Eq.~\eqref{eq:wave} on exact renderings}
\label{tab:check}
\begin{tabular}{lcccc}
\toprule
& \multicolumn{2}{c}{image frame} & \multicolumn{2}{c}{corrected frame} \\
\cmidrule(lr){2-3}\cmidrule(lr){4-5}
Rates & geometric & wave & geometric & wave \\
\midrule
$\dot\rho$ only & $4.9\times10^{-10}$ & $4.9\times10^{-10}$ & $4.9\times10^{-10}$ & $4.9\times10^{-10}$ \\
$\dot s$ only & $7.4\times10^{-10}$ & $7.4\times10^{-10}$ & $1.1\times10^{-8}$ & $1.1\times10^{-8}$ \\
$\dot X,\dot Y$ only & $1.8\times10^{-8}$ & $1.8\times10^{-8}$ & $1.8\times10^{-8}$ & $1.8\times10^{-8}$ \\
$\dot Z$ only & $1.3\times10^{-3}$ & $2.7\times10^{-9}$ & $1.3\times10^{-3}$ & $2.7\times10^{-9}$ \\
$\dot A$ only & $1.15$ & $1.1\times10^{-9}$ & $1.15$ & $1.1\times10^{-9}$ \\
$\dot\lambda$ only & $1.00$ & $8.1\times10^{-10}$ & $1.00$ & $8.1\times10^{-10}$ \\
all rates & $3.5\times10^{-4}$ & $2.4\times10^{-8}$ & $3.4\times10^{-4}$ & $2.6\times10^{-8}$ \\
\bottomrule
\end{tabular}
\tablenote{Geometric: without the diffraction term; wave: with it. Corrected frame: the magnification-corrected frame of Eq.~\eqref{eq:wave_corr}.}
\end{table}

\section{Coupled optical differentiation in wave optics}
\label{sec:cod}

COD takes depth from the ratio $q=A\,I_A/I_\rho$ of the aperture and focus derivatives, which in geometric optics equals
$-\psi$ for any aperture profile~\cite{luo2025cod}. The aperture derivative now contains the diffraction term.

\begin{proposition}[COD with a Gaussian pupil]
\label{prop:cod}
Under Lemma~\ref{lem:gauss} with $\Gamma=I$, exactly and at every frequency,
\begin{equation}
  q\equiv A\,\frac{I_A}{I_\rho}=-\psi+\frac{\psi_d^2}{\psi},\qquad \psi_d=\frac{\lambda}{4\pi A^2}.
  \label{eq:cod_gauss}
\end{equation}
\end{proposition}
\begin{proof}
With $I=G_C*P$ and $C=(\sigma^2+\sigma_d^2)I$, the heat equation gives $I_t=\tfrac12\dot V\Lap I$ with
$V=\sigma^2+\sigma_d^2$. The rates are $\partial_\rho\sigma=-As$, $\partial_A\sigma=\sigma/A$ and
$\partial_A\sigma_d=-\sigma_d/A$, so $I_\rho=-As\,\sigma\Lap I$ and $I_A=((\sigma^2-\sigma_d^2)/A)\Lap I$. Their ratio
is~\eqref{eq:cod_gauss}.
\end{proof}

The proposition has several consequences.
\begin{enumerate}
  \item The geometric COD estimate is $1/\hat Z=1/Z-\psi_d^2/\psi$, a bias with a pole at the focus distance.
  \item $q(\psi)=q(-\psi_d^2/\psi)$, so the map from depth to $q$ is two-to-one: every depth with $|\psi|<\psi_d$ pairs
  with a depth with $|\psi|>\psi_d$ on the other side of focus, and the depth map folds near focus.
  \item The fold can be undone: $\psi=\big(-q\pm\sqrt{q^2+4\psi_d^2}\big)/2$, with the branch given by
  $\operatorname{sign}\psi=-\operatorname{sign}(I_\rho/\Lap I)$, or by restricting the working range to $|\psi|>\psi_d$.
  \item At focus $I_A=-(\sigma_d^2/A)\Lap I$: diffraction turns the aperture derivative into an optical Laplacian.
  \item For an anisotropic Gaussian, $I_A=(\sigma^2/A)\tr(\Gamma\nabla\nabla^\top)I-(\sigma_d^2/A)\tr(\Gamma^{-1}
  \nabla\nabla^\top)I$, so $q$ depends on the texture unless $\Gamma\propto I$.
\end{enumerate}
Table~\ref{tab:cod_gauss} confirms~\eqref{eq:cod_gauss} with exact OTF derivatives.

\begin{table}[t]
\centering
\caption{COD ratio for a Gaussian pupil: prediction against exact OTF derivatives}
\label{tab:cod_gauss}
\begin{tabular}{rcc}
\toprule
$\psi$ [D] & predicted $q$ & numerical $q$ \\
\midrule
$-0.30$ & $0.19783$ & $0.19777$ to $0.19779$ \\
$-0.05$ & $-0.56299$ & $-0.56324$ to $-0.56307$ \\
$0.02$ & $1.51248$ & $1.51266$ to $1.51308$ \\
$0.05$ & $0.56299$ & $0.56307$ to $0.56324$ \\
$0.10$ & $0.20650$ & $0.20654$ to $0.20664$ \\
$0.30$ & $-0.19783$ & $-0.19779$ to $-0.19777$ \\
$1.00$ & $-0.96935$ & $-0.96937$ to $-0.96924$ \\
\bottomrule
\end{tabular}
\tablenote{$A=0.5$\,mm, $\psi_d^2=0.03065$\,D$^2$. Numerical values span displacements $a/(2A)=0.04$ to 0.4; the prediction is Eq.~\eqref{eq:cod_gauss}.}
\end{table}

\paragraph{Circular aperture.}
For a disc the ratio depends on frequency. Near focus, $\log H(\vf;\psi)=\log H(\vf;0)-2\pi^2\psi^2(a/\lambda)^2M(a)
+O(\psi^4)$, where $M(a)$ is the second moment along $\va$ of the normalized overlap, and for a disc of radius $R$,
$M=\tfrac{R^2}{4}\big[1-\tfrac{10}{3\pi}\tfrac aR+O((a/R)^2)\big]$~\cite{luo2026dfdd}.

\begin{proposition}[COD with a circular aperture, leading order]
\label{prop:cod_disc}
For a disc of radius $R$, small $a/R$ (equivalently $f/f_c=a/2R$) and small $\psi$,
\begin{equation}
  q(\vf)=-\psi\Big[1+\frac{5}{3\pi}\frac aR+O\big((a/R)^2\big)\Big]+\frac{D(a)}{\psi},\qquad
  D(a)=\frac{2\lambda^2}{\pi^3R^3a}\Big[1+\frac{16}{3\pi}\frac aR+O\big((a/R)^2\big)\Big].
\end{equation}
\end{proposition}
\begin{proof}
The in-focus OTF is the Chinese-hat function of $x=a/2R$, $H(\vf;0)=\tfrac2\pi\big(\arccos x-x\sqrt{1-x^2}\big)$, so
$R\,\partial_R\log H(\vf;0)=\tfrac4\pi x\sqrt{1-x^2}/H(\vf;0)=\tfrac2\pi\tfrac aR\big[1+\tfrac2\pi\tfrac aR+
O((a/R)^2)\big]$. With $M=\tfrac{R^2}{4}m(a/R)$ and $m(x)=1-\tfrac{10}{3\pi}x+O(x^2)$,
$R\,\partial_RM/(2M)=1+\tfrac{5}{3\pi}\tfrac aR+O((a/R)^2)$. Dividing $R\,\partial_R\log H$ by
$\partial_\rho\log H=4\pi^2\psi(a/\lambda)^2M$ gives the result; the bracket of $D(a)$ collects the second term of
$R\,\partial_R\log H(\vf;0)$ and the factor $1/m=1+\tfrac{10}{3\pi}\tfrac aR$, and $\tfrac2\pi+\tfrac{10}{3\pi}=
\tfrac{16}{3\pi}$.
\end{proof}

Table~\ref{tab:cod_disc} compares this with the exact ratio, computed by one-dimensional quadrature of the overlap
integral with derivatives in closed form. The leading-order formula is within 2\% of the exact ratio up to $a/R=0.1$
($f/f_c=0.05$), except near the sign change of $q$, and errs by 3--14\% at $a/R=0.2$; beyond that the exact values are
needed. Without the bracket of $D(a)$ it would err by 7\% already at $a/R=0.04$. The geometric estimate $\hat\psi=-q$ is off by an order of magnitude
near focus at low frequencies, and it depends on frequency even at $\psi=1$\,D. The invariance of COD to the aperture
shape is therefore a property of geometric optics: in wave optics it survives exactly only for Gaussian pupils. For a
hard aperture the ratio at a pixel is a texture-weighted average of $q(\vf)$, with its largest error at low
frequencies, whereas the error of the heat equation is largest at high frequencies~\cite{luo2026dfdd}.

\begin{table}[t]
\centering
\caption{COD ratio for a circular aperture: exact (quadrature) against leading order}
\label{tab:cod_disc}
\begin{tabular}{rrrcc}
\toprule
$\psi$ [D] & $a/R$ & $f/f_c$ & exact $q$ & leading order \\
\midrule
0.10 & 0.04 & 0.020 & $5.121$ & $5.107$ \\
0.10 & 0.10 & 0.050 & $2.213$ & $2.177$ \\
0.10 & 0.20 & 0.100 & $1.275$ & $1.196$ \\
0.10 & 0.40 & 0.200 & $0.891$ & $0.698$ \\
0.30 & 0.04 & 0.020 & $1.433$ & $1.430$ \\
0.30 & 0.10 & 0.050 & $0.453$ & $0.445$ \\
0.30 & 0.20 & 0.100 & $0.121$ & $0.104$ \\
0.30 & 0.40 & 0.200 & $-0.054$ & $-0.091$ \\
1.00 & 0.04 & 0.020 & $-0.504$ & $-0.500$ \\
1.00 & 0.10 & 0.050 & $-0.835$ & $-0.825$ \\
1.00 & 0.20 & 0.100 & $-1.005$ & $-0.975$ \\
1.00 & 0.40 & 0.200 & $-1.246$ & $-1.130$ \\
\bottomrule
\end{tabular}
\tablenote{$R=1$\,mm, $\lambda=550$\,nm. Exact: quadrature of the overlap integral with closed-form derivatives
(\texttt{theory/cod\_disc\_quad.py}); leading order from Proposition~\ref{prop:cod_disc}. The geometric estimate would
report $\hat\psi=-q$.}
\end{table}

\section{Only Gaussian pupils make defocus a diffusion}
\label{sec:unique}

A phase mask or an apodizer changes the pupil, so one may hope that another pupil, for example a hard aperture with a
phase mask, also makes the heat equation exact. We allow the most general local form, in which the defocus derivative
of the image is a combination of the image and its first and second derivatives.

\begin{definition}
Defocus is \emph{diffusive} for a pupil $p$ if, for all $\vf$ and $\psi$, $H(\vf;\psi)\ne0$ and
\begin{equation}
  \partial_\psi\log H(\vf;\psi)=-4\pi^2s^2\,\vf^\top B(\psi)\vf+i2\pi s\,\bm b(\psi)\cdot\vf ,
  \label{eq:closure}
\end{equation}
that is, if every image obeys the diffusion equation $\partial_\psi I=s^2\tr(B\nabla\nabla^\top)I+s\,\bm b\cdot\nabla
I$, with a diffusion matrix $B$ and a drift $\bm b$.
\end{definition}

\noindent Because $H(-\vf)=\overline{H(\vf)}$ and $H(0)=1$, there is no constant term, $B(\psi)$ is real symmetric and
$\bm b(\psi)$ is real. The drift lets the image shift with defocus, as in dual-pixel sensors.

\begin{theorem}
\label{thm:unique}
Assume (R1) that $|p|^2$ has finite moments of all orders, and (R2) that $p=|p|e^{i\phi}$ with $\phi\in C^3(\mathbb R^2)$,
whose derivatives up to order three grow at most polynomially. Then defocus is diffusive for $p$ if and only if
\begin{equation}
  p(\vu)=C\exp\!\big(-(\vu-\vu_0)^\top Q(\vu-\vu_0)+i\,\vu^\top\Theta\vu+i\,\bm\eta\cdot\vu\big),
  \label{eq:gauss_family}
\end{equation}
with a constant $C$, $Q$ positive definite, $\Theta$ symmetric and $\bm\eta$ a constant vector: the amplitude is
Gaussian, possibly anisotropic and off-center, and the phase is at most quadratic.
\end{theorem}

We write $\mathcal B(\psi)=\int_0^\psi B$ and $\mathfrak b(\psi)=\int_0^\psi\bm b$.

\paragraph{Proof of ``if''.}
The overlap of~\eqref{eq:gauss_family} is a Gaussian in $\vu$ whose shape does not depend on $\va$, multiplied by a
phase that is linear in $\vu$. Its Fourier transform at $\psi\va/\lambda$ is the exponential of a quadratic polynomial
in $\va$ for every $\psi$. A center offset $\vu_0$ adds the term $-i2\pi\psi s\,\vu_0\cdot\vf$, a lateral image shift
proportional to the defocus (Fig.~\ref{fig:kinds}c). Table~\ref{tab:closure} confirms this numerically for a Gaussian
pupil and for a Gaussian pupil with astigmatism.

\paragraph{Proof of ``only if'', step 1: amplitude.}
Fix a unit vector $\ve$ and integrate~\eqref{eq:closure} from $0$ to $\psi$:
$\log H(\vf;\psi)=\log H(\vf;0)-4\pi^2s^2\vf^\top\mathcal B(\psi)\vf+i2\pi s\,\mathfrak b(\psi)\cdot\vf$. Put
$\vf=\varepsilon\ve$ and $\psi=\nu/(s\varepsilon)$ with $\nu$ fixed, and let $\varepsilon\to0$. Then
$\va=\lambda s\varepsilon\ve\to0$ while $\psi s\vf=\nu\ve$ stays fixed, so by dominated convergence the left-hand side
tends to $\log\Phi(\nu\ve)$, where $\Phi$ is the characteristic function of $|p|^2/\|p\|^2$. On the right,
$\log H(\varepsilon\ve;0)\to0$, and the limits $L(\nu)=\lim\varepsilon^2\,\ve^\top\mathcal B(\nu/(s\varepsilon))\ve$
and $\ell(\nu)=\lim\varepsilon\,\mathfrak b(\nu/(s\varepsilon))\cdot\ve$ must exist. Replacing $\varepsilon$ by
$\varepsilon/\tau$ shows that $L$ is homogeneous of degree 2 and $\ell$ of degree 1. With
$\Phi(-\nu\ve)=\overline{\Phi(\nu\ve)}$ this gives $\log\Phi(\nu\ve)=-c_{\ve}\nu^2+i\,m_{\ve}\nu$ for all $\nu$. Every
one-dimensional projection of $|p|^2$ is therefore Gaussian, and so is $|p|^2$, by the Cram\'er--Wold theorem.

\paragraph{Step 2: phase.}
After step 1, $|p(\vu)|\propto\exp(-(\vu-\vu_0)^\top Q(\vu-\vu_0))$, so $p\ne0$ everywhere and $g=|p|^2/\|p\|^2$ is the
density of $U\sim\mathcal N(\vu_0,\Sigma)$ with $\Sigma=(4Q)^{-1}$. With $\phi$ from (R2),
\begin{equation}
  O_{\va}(\vu)=e^{-\va^\top Q\va/2}\,g(\vu)\,e^{i\Delta_{\va}(\vu)},\qquad
  \Delta_{\va}(\vu)=\phi(\vu-\tfrac{\va}{2})-\phi(\vu+\tfrac{\va}{2}).
\end{equation}

\emph{(a) Exact structure along one direction.}
Fix a unit vector $\ve$, put $\va=\varepsilon\ve$, and let
$K(\varepsilon,\kappa)=\log\int g\,e^{i\Delta_{\varepsilon\ve}}\,e^{-i2\pi\kappa\,\ve\cdot\vu}\dd\vu$. By (R1) and
(R2), $K$ is $C^2$ near $\{0\}\times\mathbb R$, and
$K(0,\kappa)=\log\hat g(\kappa\ve)=-i2\pi\kappa\,\ve\cdot\vu_0-2\pi^2V_{\ve}\kappa^2$ with $V_{\ve}=\ve^\top\Sigma\ve$.
$H$ is the Fourier transform of $O_{\va}$ at $\psi\va/\lambda=\kappa\ve$, with $\kappa=\varepsilon\psi/\lambda$, and
$s|\vf|=\varepsilon/\lambda$. The integrated equation therefore reads
\begin{equation}
  K(\varepsilon,\kappa)-K(\varepsilon,0)=
  -4\pi^2\frac{\varepsilon^2}{\lambda^2}\,\mathcal B_{\ve}\!\Big(\frac{\lambda\kappa}{\varepsilon}\Big)
  +i2\pi\frac{\varepsilon}{\lambda}\,\mathfrak b_{\ve}\!\Big(\frac{\lambda\kappa}{\varepsilon}\Big),
  \qquad \mathcal B_{\ve}=\ve^\top\mathcal B\ve,\ \ \mathfrak b_{\ve}=\mathfrak b\cdot\ve .
\end{equation}
Differentiating twice in $\kappa$ and evaluating at $\kappa=\varepsilon\psi/\lambda$ gives
\begin{equation*}
  \partial_\kappa^2K(\varepsilon,\varepsilon\psi/\lambda)=-4\pi^2\mathcal B_{\ve}''(\psi)
  +i2\pi(\lambda/\varepsilon)\,\mathfrak b_{\ve}''(\psi).
\end{equation*}
As $\varepsilon\to0$ the left-hand side tends to $-4\pi^2V_{\ve}$. The real part gives
$\mathcal B_{\ve}''\equiv V_{\ve}$, and the imaginary part would diverge unless $\mathfrak b_{\ve}''\equiv0$. Hence,
for all $\varepsilon$ and $\kappa$,
\begin{equation}
  K(\varepsilon,\kappa)-K(\varepsilon,0)=-4\pi^2\frac{\varepsilon}{\lambda}\,\eta_{\ve}\,\kappa-2\pi^2V_{\ve}\kappa^2
  +i2\pi\mu_{\ve}\kappa,\qquad \eta_{\ve}=\mathcal B_{\ve}'(0),\ \ \mu_{\ve}=\mathfrak b_{\ve}'(0).
  \label{eq:K_exact}
\end{equation}

\emph{(b) First order in $\varepsilon$.}
Differentiate~\eqref{eq:K_exact} in $\varepsilon$ at $\varepsilon=0$, which (R2) allows. Since
$\partial_\varepsilon\Delta_{\varepsilon\ve}|_{\varepsilon=0}=-\partial_{\ve}\phi$,
\begin{equation}
  F_{\ve}(\kappa)-F_{\ve}(0)=-i\,\frac{4\pi^2\eta_{\ve}}{\lambda}\,\kappa,\qquad
  F_{\ve}(\kappa)=\frac{\int g\,\partial_{\ve}\phi\;e^{-i2\pi\kappa\,\ve\cdot\vu}\dd\vu}{\int g\,e^{-i2\pi\kappa\,\ve\cdot\vu}\dd\vu}.
\end{equation}
By the Fourier slice theorem, numerator and denominator are the one-dimensional Fourier transforms of the projections
$\rho_{\ve}(t)=\int_{\ve\cdot\vu=t}g\,\partial_{\ve}\phi$ and $g_{\ve}(t)=\int_{\ve\cdot\vu=t}g$. An affine $F_{\ve}$
means $\rho_{\ve}=F_{\ve}(0)\,g_{\ve}+c\,g_{\ve}'$ for a constant $c$. Since $g_{\ve}$ is Gaussian, $g_{\ve}'/g_{\ve}$ is
affine, and
\begin{equation}
  \mathbb E\big[\partial_{\ve}\phi(U)\,\big|\,\ve\cdot U=t\big]=\frac{\rho_{\ve}(t)}{g_{\ve}(t)}
  \ \text{ is affine in } t,\ \text{ for every unit vector } \ve .
  \label{eq:cond_affine}
\end{equation}

\emph{(c) Chaos expansion.}
Write $U=\vu_0+\Sigma^{1/2}W$ with $W\sim\mathcal N(0,I)$ and $\tilde\phi(\bm w)=\phi(\vu_0+\Sigma^{1/2}\bm w)$, and
expand $\tilde\phi=\sum_n\tilde\phi_n$ in Wiener chaos, that is, in two-variable Hermite polynomials ($\tilde\phi\in L^2$
by (R2)). For a unit vector $\ve$ let $\hat{\bm v}=\Sigma^{1/2}\ve/|\Sigma^{1/2}\ve|$ and $\bm d=\Sigma^{-1/2}\ve$. Then
$\ve\cdot U$ is an affine function of $\hat{\bm v}\cdot W$, and $\partial_{\ve}\phi(U)=\bm d\cdot\nabla\tilde\phi(W)$. In
the orthonormal frame $(\hat{\bm v},\hat{\bm v}^\perp)$,
$\tilde\phi_n=\sum_{j+k=n}c_{jk}\,\mathrm{He}_j(\hat{\bm v}\cdot W)\,\mathrm{He}_k(\hat{\bm v}^\perp\cdot W)$. With
$\mathrm{He}_j'=j\,\mathrm{He}_{j-1}$ and $\mathbb E[\mathrm{He}_k(\hat{\bm v}^\perp\cdot W)]=\delta_{k0}$,
\begin{equation}
  \mathbb E\big[\bm d\cdot\nabla\tilde\phi(W)\,\big|\,\hat{\bm v}\cdot W\big]
  =\sum_{n\ge1}(\bm d\cdot\nabla\pi_n)(\hat{\bm v})\;\mathrm{He}_{n-1}(\hat{\bm v}\cdot W),
\end{equation}
where $\pi_n(\bm w)=\sum_{j+k=n}c_{jk}(\hat{\bm v}\cdot\bm w)^j(\hat{\bm v}^\perp\cdot\bm w)^k$ is the leading
homogeneous part of $\tilde\phi_n$, a polynomial that does not depend on the frame. By~\eqref{eq:cond_affine} the
coefficients with $n\ge3$ vanish: $(\bm d\cdot\nabla\pi_n)(\hat{\bm v})=0$. By homogeneity this is
$(\Sigma^{-1}\bm v)\cdot\nabla\pi_n(\bm v)=0$ for every $\bm v\in\mathbb R^2$. So $\pi_n$ is constant along the
trajectories of $\dot{\bm v}=\Sigma^{-1}\bm v$, all of which emerge from the origin as $t\to-\infty$. With $\pi_n(0)=0$
this gives $\pi_n\equiv0$. The leading-part map is injective on the $n$-th chaos, so $\tilde\phi_n=0$ for $n\ge3$, and
$\phi$ is a polynomial of degree at most 2. \hfill$\square$

\begin{remark}[Why every direction is needed]
The phase $(u_xu_y)^2$ satisfies~\eqref{eq:cond_affine} along both coordinate axes: for $\Sigma=I$,
$\mathbb E[2U_xU_y^2\,|\,U_x]=2U_x$. Along the diagonal it does not: with $t=(u_x+u_y)/\sqrt2$,
$\mathbb E[\partial_{\ve}\phi\,|\,\ve\cdot U=t]=t^3-t$. A proof that checked only the axes would miss it, and
Table~\ref{tab:closure} shows that, once higher orders in $a$ are included, exact diffusion fails for this phase in both
directions.
\end{remark}

\begin{remark}[What the proof uses]
Step 1 uses the limit $\va\to0$, and step 2 the first order in $\va$. With $\phi=2\pi W/\lambda$ for a wavefront error
$W$, these correspond to the geometric PSF and to the transverse ray aberration $s\nabla W$. This suggests that the same
classification holds for aberrated pupils in geometric optics, but we do not prove it here. Alexander's
theorem~\cite{alexander2019} is stated for geometric optics and nonnegative aperture profiles; Theorem~\ref{thm:unique}
covers complex pupils in wave optics, but its definition fixes the order of the local constraint at two.
\end{remark}

\paragraph{Numerical test.}
Table~\ref{tab:closure} evaluates the diffusion coefficient $c(\va;\psi)=-\partial_\psi\log H/(4\pi^2\psi(a/\lambda)^2)$
along one direction for several pupils, relative to its value at the smallest displacement. Exact diffusion gives 1 at
every displacement. Only the Gaussian pupils, with and without astigmatism, pass. A truncated Gaussian, a
super-Gaussian amplitude, a hard disc and a Gaussian with a quartic phase all fail, the last one along both directions.

\begin{table}[t]
\centering
\caption{Test of exact diffusion for several pupils}
\label{tab:closure}
\begin{tabular}{lcccccc}
\toprule
$a/R$ & 0.02 & 0.05 & 0.10 & 0.20 & 0.30 & 0.40 \\
\midrule
Gaussian & 1.0000 & 1.0000 & 1.0000 & 1.0000 & 1.0000 & 1.0000 \\
Gaussian + astigmatism 0.3\,D ($x$ and $y$) & 1.0000 & 1.0000 & 1.0000 & 1.0000 & 1.0000 & 1.0000 \\
Gaussian + spherical $0.25\lambda$ & 1.0000 & 0.9478 & 0.8251 & 0.6148 & 0.4801 & 0.3930 \\
Gaussian truncated at $2A$ & 1.0000 & 0.9824 & 0.9526 & 0.8917 & 0.8290 & 0.7646 \\
super-Gaussian amplitude & 1.0000 & 0.9986 & 0.9937 & 0.9743 & 0.9429 & 0.9008 \\
hard disc & 1.0000 & 0.9686 & 0.9181 & 0.8233 & 0.7353 & 0.6529 \\
Gaussian + $(u_xu_y)^2$ phase, along $x$ & 1.0000 & 0.9799 & 0.9255 & 0.8114 & 0.7252 & 0.6637 \\
Gaussian + $(u_xu_y)^2$ phase, diagonal & 1.0000 & 0.9492 & 0.8272 & 0.6128 & 0.4740 & 0.3849 \\
\bottomrule
\end{tabular}
\tablenote{Entries: $c(\va;\psi)/c(\va_{\min};\psi)$ at $\psi=0.2$\,D along one direction; exact diffusion gives 1 everywhere. $R=1$\,mm sets the scale; Gaussian pupils have $A=R/2$.}
\end{table}

\paragraph{Consequences.}
No pupil with a hard edge, that is, with an amplitude of one inside and zero outside, makes defocus an exact
diffusion. This rules out a clear aperture with or without a phase mask, coded apertures, and phase-only metasurfaces.
A DfDD system that needs an exact constraint needs an apodized aperture with a Gaussian amplitude; apodizing filters of
this kind are made for photographic lenses. Built-in defocus and astigmatism are allowed, because they only add a
constant to $B$. An off-center Gaussian sub-aperture gives exact diffusion with a drift proportional to the defocus,
which is the mechanism of dual-pixel disparity; real half-pupils are not Gaussian, so for them the diffusion equation is
approximate. Whether defocus is an exact diffusion thus depends on the amplitude of the pupil, and no phase design can
compensate for a hard edge.

\section{Discussion}
\label{sec:discussion}

\paragraph{Diffraction in DfDD cues.}
For a Gaussian pupil, the cues that change only the focus or the sensor distance, or that move the target laterally,
are blind to diffraction: their constraints hold exactly at every wavelength and every f-number. The other cues need the
diffraction variance $\sigma_d^2=(\lambda s/4\pi A)^2$. For axial motion it is a small correction (0.13\% in our test),
but for aperture and wavelength changes it is of the same order as the signal near focus, and for COD it creates a fold
that must be resolved by the sign of the focus derivative.

\paragraph{Hard apertures.}
Theorem~\ref{thm:unique} says that every hard aperture violates the heat equation, but not how. The companion
paper~\cite{luo2026dfdd} answers this for the circular aperture: near focus the diffusion coefficient at a spatial
frequency is the second moment of the overlap in Fig.~\ref{fig:kinds}b, which narrows as the frequency grows, so the
coefficient falls, to first order, by the fraction $\tfrac{20}{3\pi}f/f_c$. On released data of a DfDD prototype the
measured coefficient falls with frequency at three iris diameters, along one curve in $f/f_c$. A test fixed in advance
confirmed a fall on new smartphone sweeps, at 95\,cycles/mm only and steeper than a circular aperture predicts. On a
DSLR, the turnovers and sign changes of narrow frequency bands exclude a disc of any one radius in geometric optics,
while a test of the fall in the band slopes, also fixed in advance, was refuted.

\paragraph{Open problems.}
The theory is scalar and paraxial, for quasi-monochromatic and spatially incoherent light. Local constraints of order
higher than two in wave optics are not classified, and neither is the geometric-optics counterpart of
Theorem~\ref{thm:unique} for aberrated pupils, which the proof suggests.

\section{Conclusion}
\label{sec:conclusion}

Writing the optical transfer function as an integral over the overlap of two displaced pupils turns the question of
exact differential-defocus constraints into a question about the shape of that overlap. For a Gaussian pupil the shape
does not change, the PSF stays Gaussian, and every DfDD constraint holds in closed form, with a diffraction term for the
cues that change the aperture, the wavelength or the magnification. For every other pupil the shape changes, and
defocus is not an exact diffusion: the Gaussian amplitude, possibly anisotropic and off-center, with at most a quadratic
phase, is the only exact case.

\bmsection{Disclosures} The author declares no conflicts of interest.

\bmsection{Data availability} Data underlying the results presented in this paper are not publicly available at this
time but may be obtained from the author upon reasonable request.

\end{document}